\documentclass[aps,prl,twocolumn,showpacs,superscriptaddress,longbibliography,floatfix]{revtex4-2}

\usepackage{amsmath, amsfonts, amssymb, bm}
\usepackage{mathtools}
\usepackage[inline,shortlabels]{enumitem}
\usepackage{graphicx}
\usepackage{comment}
\usepackage{xcolor}
\usepackage[breaklinks,hypertexnames=false]{hyperref}
\hypersetup{colorlinks,linkcolor={magenta},citecolor={blue},urlcolor={blue}} 

\usepackage[capitalize]{cleveref}
\usepackage{multirow}

\usepackage{glossaries}
\glsdisablehyper

\newacronym{mbs}{MBS}{Majorana bound state}
\newacronym{zbp}{ZBCP}{zero-bias conductance peak}
\newacronym{zes}{qZES}{quasi zero-energy state}

\newacronym{dos}{LDOS}{local density of states}
\newacronym{bdg}{BdG}{Bogoliubov-de Gennes}

\newcommand{\e}{\mathrm{e}}

\newcommand{\mbf}[1]{\mathbf{ #1 }}

\DeclarePairedDelimiter\mean{\langle}{\rangle}

\DeclareMathOperator{\sgn}{sgn}

\begin{document}
	
\title{Confinement-induced zero-bias peaks in conventional superconductor hybrids} 

\author{Jorge Cayao} 
\affiliation{Department of Physics and Astronomy, 
	Uppsala University, Box 516, S-751 20 Uppsala, Sweden}
\affiliation{Theoretische Physik, Universit\"{a}t Duisburg-Essen and CENIDE, D-47048 Duisburg, Germany}

\author{Pablo Burset}
\affiliation{Department of Theoretical Condensed Matter Physics, Universidad Aut\'onoma de Madrid, 28049 Madrid, Spain}
	
	\date{\today}
	
\begin{abstract}
Majorana bound states in topological superconductors have been predicted to appear in the form of zero-bias conductance peaks of height $2e^{2}/h$, which represents one of the most studied signatures so far. Here, we show that quasi zero-energy states, similar to Majorana bound states, can naturally form in any superconducting hybrid junction due to confinement effects, without relation to topology. Remarkably, these topologically trivial quasi zero-energy states produce zero-bias conductance peaks, thus mimicking the most representative Majorana signature. Our results put forward confinement as an alternative mechanism to explain the ubiquitous presence of trivial zero-bias peaks and quasi zero-energy states in superconductor hybrids.
\end{abstract}

\maketitle
	

{\it Introduction.---}The realization of topological superconductivity featuring \glspl{mbs} in superconducting hybrid systems has lately been the subject of intense research due to its potential for technological applications~\cite{Fujimoto_JPSJ,Sato_2017,Aguadoreview17,lutchyn2018majorana,zhang2019next,beenakker2019search,doi:10.1146/annurev-conmatphys-031218-013618,2020Aguado}. 
The most promising approach involves semiconductor nanowires with strong spin-orbit coupling and proximity-induced conventional  superconductivity~\cite{DasSarma_2010,VonOppen_2010}. 
Here, an external magnetic field drives such systems into a topological phase where \glspl{mbs} emerge at zero energy and localize at each end of the wire. These properties enable \glspl{mbs} to form the basis for qubit proposals robust against local perturbations~\cite{kitaev,PhysRevLett.86.268,RevModPhys.80.1083,Sarma:16,alicea2011non,PhysRevX.6.031016}, highlighting the importance of topological superconductivity in condensed matter physics. 

The detection of \glspl{mbs} has been mainly pursued by exploiting their zero-energy nature. Indeed, tunneling from a normal metal into a \gls{mbs} has been predicted to result in a \gls{zbp} of height $2e^{2}/h$~\cite{Kashiwaya_RPP,PhysRevLett.98.237002,PhysRevLett.103.237001,PhysRevB.82.180516,Lu_2016,Burset_2017}. Subsequently, many experiments reported \glspl{zbp} and interpreted them as strong evidence of \glspl{mbs}~\cite{Mourik:S12,Higginbotham,Deng16,Albrecht16,zhang16,Suominen17,Nichele17,zhang18,gul2018ballistic}. Despite the significant experimental progress, the Majorana origin of \glspl{zbp} has been recently questioned. In great part because several works have reported \glspl{zbp} due to \glspl{zes} at finite magnetic fields but well below the topological phase and, hence, not tied to topology~\cite{PhysRevB.86.100503,PhysRevB.86.180503,PhysRevB.91.024514,JorgeEPs,StickDas17,Ptok2017Controlling,Fer18,PhysRevB.98.245407,PhysRevLett.123.107703,PhysRevB.100.155429,PhysRevLett.123.217003,10.21468/SciPostPhys.7.5.061,avila2019non,PhysRevResearch.2.013377,PhysRevLett.125.017701,PhysRevLett.125.116803,Olesia2020,yu2020non,valentini2020nontopological,prada2019andreev,Ziani_2020}. 
In this regard, recent theoretical efforts have suggested interesting detection protocols of \glspl{mbs}~\cite{PhysRevB.97.165302,PhysRevB.97.214502,PhysRevB.97.161401,PhysRevB.99.155159,PhysRevB.100.241408,PhysRevLett.123.117001,PhysRevB.101.014512,PhysRevB.102.045111,Kashuba_2017,PhysRevB.102.045303,doi:10.1002/qute.201900110,pan2020threeterminal,liu2020topological,PhysRevLett.124.036801,ricco2020topological,thamm2020transmission,chen2020nonabelian,cayao2020distinguishing,Fleckenstein_2021,arXiv:2105.05925}, but the emergence of trivial \glspl{zes} still seems puzzling in conductance measurements. 

Given the complex experimental setups involved in all Majorana platforms, it is fair to say that it still remains unclear if the emergence of trivial \glspl{zes}, and associated \glspl{zbp}, is due to the interplay between superconductivity and magnetism, the intrinsic inhomogeneities of superconducting heterostructures, or both. Understanding the mechanisms behind these non-topological states can help to interpret \glspl{zbp} in experiments, rule out a possible Majorana origin, and envisage routes for mitigating the emergence of the unwanted \glspl{zes}. 
\begin{figure}[!t]
	\includegraphics[width=0.95\columnwidth]{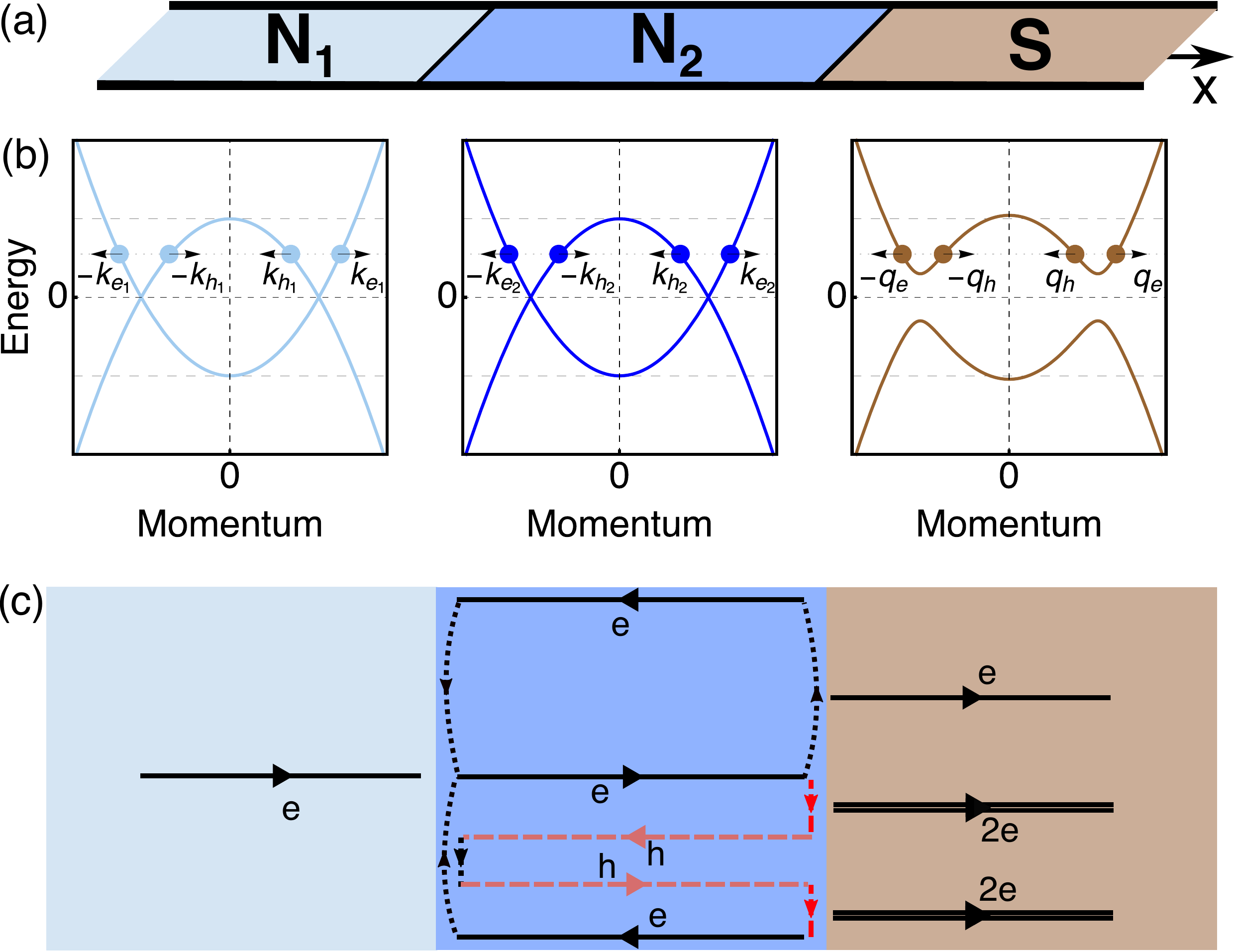}
	\caption{\label{fig:sketch} 
		(a) Sketch of a one-dimensional N$_{1}$N$_{2}$S junction. 
		(b) Dispersion relation on each region at the same chemical potential. The arrows represent the velocity direction of electrons and holes with wavevectors $k_{e,h}$ in N$_{1,2}$ and $q_{e,h}$ in S. 
		(c) Quasi-bound states in N$_{2}$ involving only normal reflections (top) or both normal and Andreev processes (bottom). 
	}
\end{figure}

In this work we demonstrate that \glspl{zes} can naturally emerge in hybrid junctions with conventional $s$-wave superconductors just due to confinement, and thus requiring neither magnetism nor spin-orbit coupling. To illustrate this generic effect, we consider a normal metal-normal metal-superconductor (N$_{1}$N$_{2}$S) junction, as in \cref{fig:sketch}(a), where confinement in the central region (N$_{2}$) enables the formation of \glspl{zes}. Interestingly, we find that these trivial \glspl{zes} produce almost perfectly quantized \glspl{zbp} that mimic those due to \glspl{mbs}. Since most of the setups used to detect \glspl{mbs} are prone to confinement, the emergence of topologically trivial \glspl{zes}, with quantized \glspl{zbp}, is a generic and ubiquitous effect in any superconducting hybrid system. 


{\it Theoretical formulation.---}We consider a ballistic one-dimensional  junction between a normal metal and a superconductor separated by a normal region of length $L$, as  shown in \cref{fig:sketch}(a). This can be modeled by the  Bogoliubov-de Gennes (BdG) Hamiltonian given by,
\begin{equation}
	\label{Eq1}
	H=\bigg[ \frac{p^{2}}{2m} - \mu(x)\bigg] \tau_{z}+\Delta(x)\tau_{x} ,
\end{equation}  
where $p=-i\hbar\partial_{x}$ is the momentum operator, $m$ the electron effective mass, $\mu(x)=\mu_{\rm N_{1}}\Theta(-x-L)+\mu_{\rm N_{2}}\theta(x+L)\Theta(-x)+\mu_{\rm S}\Theta(x)$ represents the chemical potential profile across the junction, with $\Theta(x)$ being the Heaviside step function and $\mu_{\rm N_{1(2)}}$ and $\mu_{\rm S}$ the chemical potentials in $N_{1(2)}$ and S regions, respectively. 
Moreover, $\Delta(x)=\Delta\Theta(x)$ represents the conventional singlet $s$-wave pair potential with $\Delta\neq0$ only in S. We contrast our findings with junctions where S is a topological superconductor, which we model substituting the pair potential in \cref{Eq1} by the spin-triplet $p$-wave $\Delta(x)=\sgn(p)\Delta\Theta(x)$~\cite{PhysRevB.100.115433}. 

Diagonalizing the Hamiltonian in \cref{Eq1}, we obtain the energy-momentum dispersion in each region presented in \cref{fig:sketch}(b), with wavevectors in the N regions given by $k_{e_{i},h_{i}}= k_{\rm N_{i}} \sqrt{1\pm \omega/\mu_{\rm N_{i}}}$, with $k_{\rm N_{i}}=\sqrt{2m\mu_{\rm N_{i}}/\hbar^{2}}$ and $i=1,2$. 
In the S region, we obtain $q_{e,h}=k_{\rm S}\sqrt{1\pm\sqrt{\omega^{2}-\Delta^{2}}/\mu_{\rm S}}$, with $k_{\rm S}=\sqrt{2m\mu_{\rm S}/\hbar^{2}}$. 
These wavevectors characterize the right and left moving electrons and holes (electron-like and hole-like quasiparticles) in the normal (superconducting) regions, indicated by filled circles with horizontal arrows in \cref{fig:sketch}(b). 
Next, we use the scattering states associated to these quasiparticles to show how confinement in the middle region enables the formation of \glspl{zes}.


{\it Confinement in normal-state junctions.---}To understand the origin and impact of confinement in hybrid junctions modeled by \cref{Eq1}, we first inspect the role of the intermediate N$_{2}$ region on transport across the junction when S is in its normal state (i.e., $\Delta=0$). For this purpose, we calculate the normal transmission across the NN$_2$N junction by matching the scattering states at the system interfaces, obtaining
\begin{equation}
	\label{EqTN}
	T_{\rm N}=\frac{2e^{2}/h}{1+\big[\frac{k_{e}}{2k_{e_{2}}}+\frac{k_{e_{2}}}{2k_{e}}\big]^{2}-\big[\frac{k_{e}}{2k_{e_{2}}}-\frac{k_{e_{2}}}{2k_{e}}\big]^{2}{\rm cos}(2k_{e_{2}}L)} ,
\end{equation}
where, without loss of generality, we assumed that the outer regions have the same chemical potential $\mu$ (i.e., $k_{e1}=q_{e}\equiv k_{e}$), while N$_{2}$ has $\mu_{\rm N_{2}}$ and length $L$. For a detailed derivation of \cref{EqTN}, see Supplementary Material (SM)~\cite{SM}. The normal transmission $T_{\rm N}$, \cref{EqTN}, describes the possibility of an incident electron to be transmitted through the junction after experiencing several normal reflections inside the middle N$_{2}$ region. 
The effect of N$_{2}$ is captured in the cosine term in the denominator of \cref{EqTN}, which signals the appearance of confinement and enables the formation of discrete energy levels whose number depends on the length of N$_{2}$, similarly to a Fabry-Perot cavity. Consequently, there is a resonant transmission $T_{\rm N}/(e^{2}/h)=1$ either in the absence of N$_{2}$, i.e., for $L=0$ which leads to $\cos(2k_{e_{2}}L)=1$, or when the wave vectors of the three regions are the same, $k_{e}=k_{e_{2}}$, see \cref{fig:sketch}(b). For a finite length N$_{2}$ region, with a chemical potential different than the outer regions, the resonant condition is $k_{e_{2}}L=n\pi$, with $n$ and integer. 

\begin{figure*}[!ht]
	\begin{minipage}[t]{\linewidth}
		\centering
		\includegraphics[width=.99\textwidth]{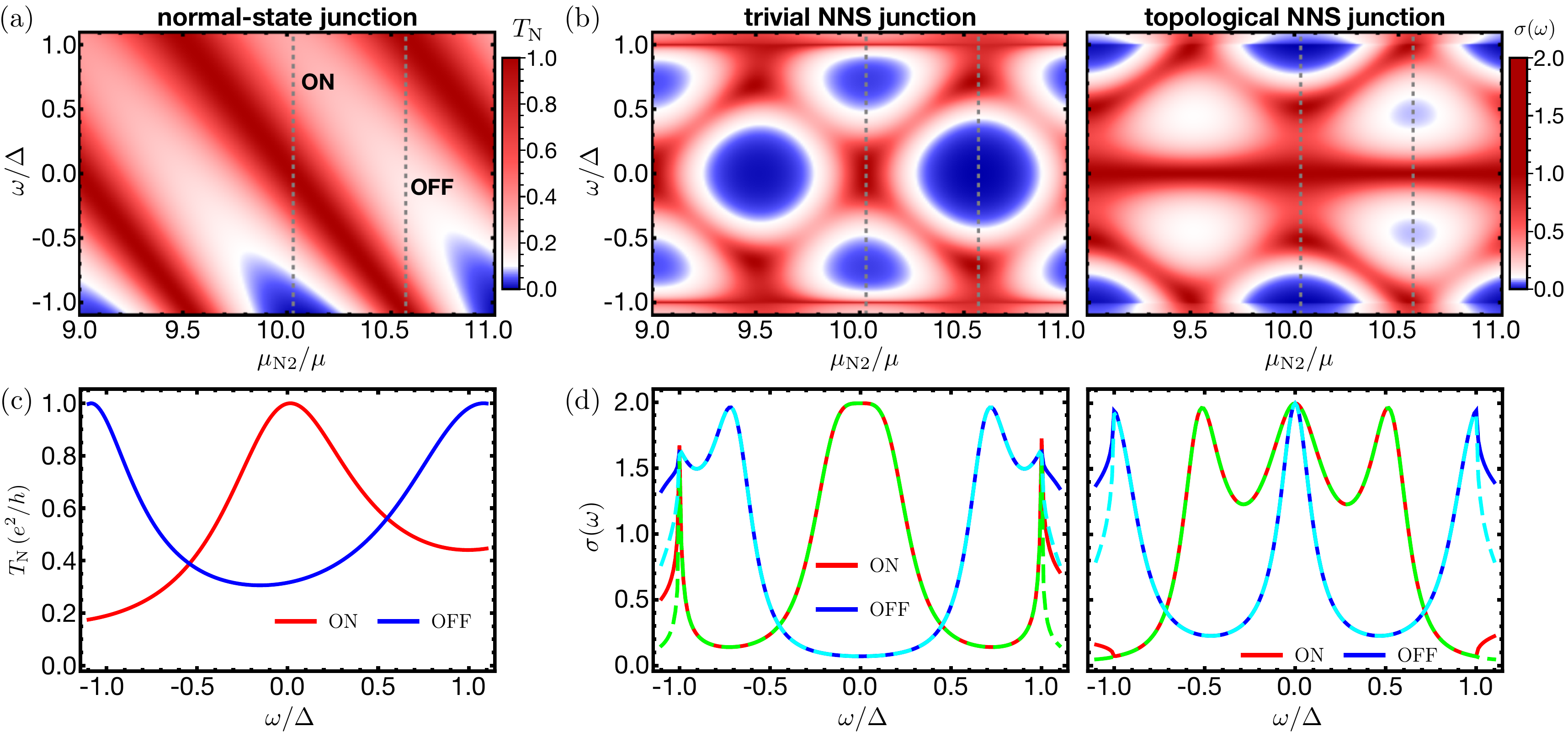} 
		\caption{
			(a,b) Map of the conductance as a function of energy and the chemical potential of the intermediate region N$_{2}$ for a junction in the normal state (a) or with a trivial and topological superconductor (b). 
			(c,d) Conductance on and off-resonance when the junction is in the normal state (c) or has a trivial or a topological superconductor (d). The values of $\mu_{\text{N}2}$ for each case are plotted as dashed gray lines in the maps, labeled ON and OFF for on- and off-resonance, respectively. For (d), we plot $2|r_{eh}|^2$ using green (cyan) dashed lines for on (off) resonance. 
			In all cases, $k_\text{F}L=3\pi/2$, $\mu_{N1}=\mu_S\equiv\mu=2\Delta$. }
		\label{fig:cond}
	\end{minipage}
\end{figure*}

To visualize this behavior, we plot $T_{\rm N}$ in \cref{fig:cond}(a) as a function of the chemical potential $\mu_{\text{N}2}$. This allows us to identify the conditions for transport \emph{on} and \emph{off} resonance when the transmission is either maximum [$T_{\rm N}/(e^{2}/h)=1$] or minimum. \emph{On resonance}, we obtain a \gls{zbp} with periodicity determined by the length $L$, cf. \cref{EqTN}. 
Interestingly, $T_{\rm N}$ can be resonant at exactly zero energy ($\omega=0$), that is, solely as a result of confinement from the middle region due to the finite length and different chemical potential, see dashed line in \cref{fig:cond}(a); see also \cref{fig:cond}(b). 
As we show next, this rather general result is behind the formation of trivial Andreev \glspl{zes} in superconducting junctions, making this effect ubiquitous in superconducting heterostructures~\cite{Klapwijk_2014}. 

{\it Confinement in superconducting junctions.---}We now analyze the consequences of confinement in transport across trivial and topological N$_{1}$N$_{2}$S junctions. 
In addition to normal reflections at both interfaces, Andreev reflections also take place at the N$_{2}$S interface, where incident electrons from N$_{2}$ are reflected back as holes~\cite{Klapwijk2004}. 
Owing to these Andreev reflections the discrete energy levels of the middle region, discussed in previous section, become coherent superpositions of electrons and holes. 
As a result, Andreev quasi-bound states are formed with properties that depend on the system parameters, as schematically shown by the bottom process of \cref{fig:sketch}(c). This occurs for both trivial and topological junctions. 
To analyze the impact of this phenomenon on transport properties, we inspect the normalized conductance~\cite{BTK}
\begin{equation}
	\label{EqdIdV}
	\sigma(\omega)=\frac{2e^2}{h}T_\text{N}^{-1} (1-|r_{ee}(\omega)|^{2}+|r_{eh}(\omega)|^{2})\,,
\end{equation}
where $T_\text{N}$ is given by \cref{EqTN}, and $r_{ee}$ and $r_{eh}$ represent the normal and Andreev reflection amplitudes, respectively. 
These amplitudes are obtained by matching the scattering states of \cref{Eq1} at the interfaces of the N$_{1}$N$_{2}$S junction~\cite{SM}. 
\Cref{EqdIdV} characterizes transport in both trivial and topological junctions, and we now discuss how it is affected by confinement effects. 

In \cref{fig:cond}(b) we map the conductance $\sigma$ as a function of the energy $\omega$ and the chemical potential of the middle region $\mu_{\rm N_{2}}$ for a trivial and a topological (right panel) N$_{1}$N$_{2}$S junction. 
For both trivial and topological junctions, the conductance develops resonances with maximum values (dark red areas) that double those obtained when S is in its normal state, see \cref{fig:cond}(a). 
Here, the most important feature is that the conductance for a trivial N$_{1}$N$_{2}$S junction exhibits a \gls{zbp} for exactly the same parameters that result in a resonant normal-state transmission, see \cref{fig:cond}. 
Note that the appearance of this \gls{zbp} can be tuned by the chemical potential of the middle region $\mu_{\text{N}2}$. 
We have verified that impurity scattering at the interfaces narrows the width of this \gls{zbp}. 
The presence or absence of a \gls{zbp} directly corresponds to the \emph{on-} or \emph{off-}resonance regimes of the normal-state conductance, marked by vertical dotted lines in \cref{fig:cond}(a,b). 
By contrast, the \gls{zbp} in topological junctions remains robust for any value of $\mu_{\text{N}2}$. On resonance, however, it can be challenging to identify whether the \gls{zbp} is of trivial or topological nature as in both cases it exhibits a very similar behavior. 

On resonance, the \gls{zbp} for a trivial superconductor is properly quantized as twice the normal state transmission, compare \cref{fig:cond}(c) and \cref{fig:cond}(d). Such quantization quickly disappears if $\mu_{\text{N}2}$ is tuned out of resonance. 
Moreover, the \gls{zbp} is mainly due to Andreev processes fulfilling $\sigma(|\omega|<\Delta)=2|r_{eh}|^2$, with $|r_{eh}(\omega=0)|^2\rightarrow1$, see green and cyan dashed lines in \cref{fig:cond}(d). 
By contrast, the topological superconductor features a \gls{zbp} both on and off resonance, with a perfect zero energy Andreev reflection $|r_{eh}(\omega=0)|^2=1$. 
While topological junctions exhibit robust unitarity of Andreev reflection, confinement in trivial junctions can only approximately accommodate regimes of unitarity. 
These two situations are, however, difficult to distinguish by naked eye. 

The findings discussed above thus show that, on-resonance, the \glspl{zbp} for both trivial and topological junctions feature a very similar behavior. We stress again that the \gls{zbp} for trivial junctions arises solely due to confinement effects of the central region, since we have not considered any magnetic order and the pair potential does not include \glspl{mbs}. Consequently, a \gls{zbp} is not a definitive indicator of \glspl{mbs} or topological superconductivity as it can easily appear in ballistic heterostructures without magnetic order or topology. 


{\it Real space zero-energy LDOS.---}To better understand the behavior of the trivial \gls{zes}, we now study the spatial dependence of the \gls{dos} $\rho$ at zero-energy. 
The \gls{dos} is obtained from the retarded Green's function $G^{r}(x,x',\omega)$ associated to the BdG Hamiltonian in \cref{Eq1}. To find $G^{r}(x,x',\omega)$ we follow a scattering Green's function approach~\cite{SM} commonly used for superconducting junctions~\cite{McMillan_1968,Furusaki_1991,Kashiwaya_RPP,Herrera_2010,Burset_2015,Lu_2015,Cayao_2017,Lu_2018}. The Green's function $G^{r}(x,x',\omega)$ is a $2\times2$ matrix in Nambu (particle-hole) space, and the \gls{dos} is then obtained as $\rho(\omega,x)= -{\rm Im}{\rm Tr}[G^{r}(x,x,\omega)]/\pi$. The \gls{dos} in S and N$_{2}$ exhibits a complex behavior, but in N$_{1}$ it is simply given by
\begin{equation}\label{LDOSEqs}
	\rho(\omega,x) = \frac{m}{\pi\hbar^2} {\rm Im}\Big\{ \sum\limits_{\alpha=e,h} \frac{i}{k_\alpha} \left( 1 + r_{\alpha\alpha} {\rm e}^{-2i s_\alpha k_{\alpha}x} \right) \Big\},
\end{equation}
where $s_{e,h}=\pm 1$ and $r_{\alpha\alpha}$ represent normal reflection amplitudes. Deep inside the leftmost normal region, the \gls{dos} adopts the simple form $\rho_0=m(k^{-1}_{e1}+k^{-1}_{h1})/(\pi\hbar^2)$, which we use for normalization. 

\begin{figure}[!t]
	\centering
	\includegraphics[width=0.95\columnwidth]{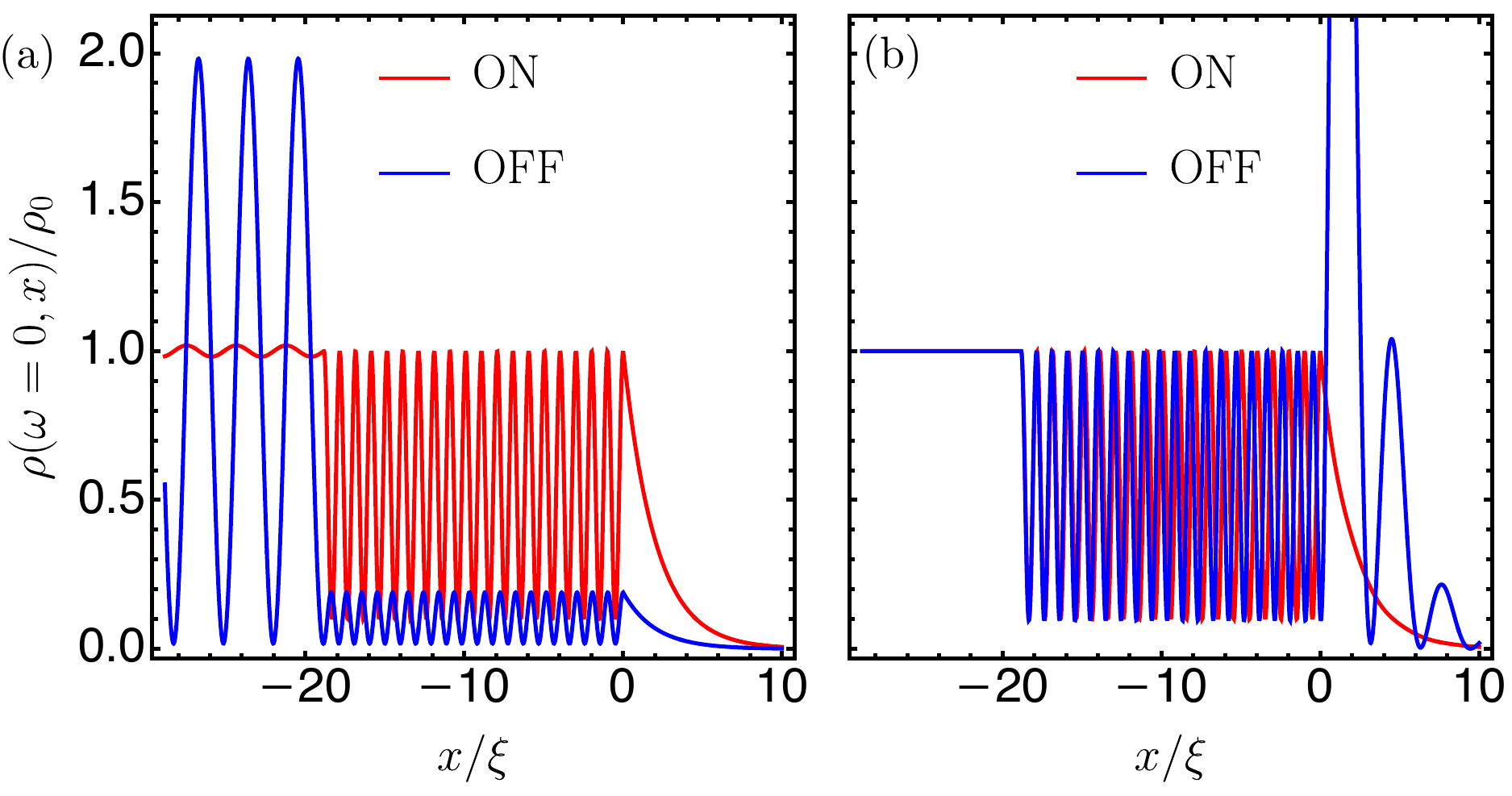}
	\caption{
		Spatial dependence of the zero-energy \gls{dos} for trivial (a) and topological (b) junctions, with the same parameters as \cref{fig:cond}, and red (blue) lines corresponding to on- (off-)resonance. 
	}
	\label{fig:LDOS}
\end{figure}

Because our interest is on the \gls{zes}, we present in \cref{fig:LDOS} the spatial dependence of the zero-energy \gls{dos} for trivial (a) and topological (b) junctions, when the chemical potential of the middle region $\mu_{\rm N_{2}}$ is set on- and off-resonance (red and blue lines, respectively). 
For topological junctions, \cref{fig:LDOS}(b), the zero-energy \gls{dos} is almost independent of the chemical potential in N$_2$, $\mu_{\rm N_{2}}$. The zero-energy \gls{dos} is perfectly flat on N$_{1}$ and oscillates with  constant amplitude in N$_2$, unaffected by variations of $\mu_{\rm N_{2}}$. Both features are a result of the perfect Andreev reflection taking place at the N$_2$-S interface for $\omega=0$, where the \gls{mbs} is located, which, in turn, also promotes a perfect transmission at N-N$_2$. 
As a result, $r_{ee}=0$ in \cref{LDOSEqs} and the zero-energy \gls{dos} in N takes exactly the value of the bulk density $\rho_0$. Hence, the robust profile of the zero-energy \gls{dos} in topological junctions can be attributed to the presence of a \gls{mbs}. 

For trivial junctions on-resonance, the magnitude and oscillations of the zero-energy \gls{dos} [red curve in \cref{fig:LDOS}(a)] are very similar to those of topological junctions, albeit there are no \gls{mbs} present in this case. 
This indicates the formation of an extended \gls{zes} in N$_2$, which is responsible for the finite \gls{dos} at zero energy at the N-N$_2$ interface and causes the \gls{zbp} in the conductance on-resonance (\cref{fig:cond}). 
However, off-resonance, the zero-energy \gls{dos} oscillations become vanishing small, as seen in the blue curve of \cref{fig:LDOS}(a). 
Moreover, the zero-energy \gls{dos} displays clear oscillations in the leftmost normal region N$_{1}$, originating from the term proportional to $r_{\alpha\alpha}$ in \cref{LDOSEqs}. While the oscillations in N$_{1}$ are present both on- and off-resonance, their amplitude is greatly suppressed on-resonance because normal reflections are finite but vanishing small, i.e., $|r_{ee}|^2\rightarrow0$. 

By the exposed above, the zero-energy \gls{dos} on-resonance for a trivial junction has the same qualitative behavior as that of a topological junction. While for trivial junctions on resonance the unitarity of the Andreev reflection is only approximately true, i.e., $|r_{eh}|^2\rightarrow1$, the presence of a \gls{mbs} in topological junctions promotes a perfect Andreev reflection $|r_{eh}|^2=1$. However, whether the Andreev reflection unitarity is approximated or exact is rather challenging to distinguish in measurements, thus posing a critical question when interpreting \glspl{zbp}. 
Consequently, discerning between the perfect Andreev reflection ($|r_{eh}|^2=1$) for a \gls{mbs} and the approximate one ($|r_{eh}|^2\lesssim1$) for a trivial \gls{zes} on resonance would require a very sensitive scanning tunneling experiment. 


{\it Finite-size effects.---}To showcase the emergence of confinement-induced \glspl{zes}, we have thus far considered a perfectly one-dimensional ballistic junction with semi-infinite outer leads. 
We now explore possible deviations from having finite-size outer leads or quasi-one dimensional junctions. 
First, we performed tight-binding simulations on finite length systems, after discretizing \cref{Eq1} on a lattice, and verified that the main findings remain robust under more realistic conditions, see Ref.~\cite{SM} for more details. 
Previously, the appearance of trivial \glspl{zes} has been confirmed in junctions with spin-orbit coupling and magnetism~\cite{PhysRevB.91.024514,cayao2020distinguishing}. However, these junctions require that the middle region is within the helical regime, and it was not clearly determined if the \glspl{zes} originate due to confinement or helicity. 
Our work thus clarifies the pivotal role of confinement in the formation of \glspl{zes} in semiconductor hybrid junctions. 

Another finite-size effect is related to the finite cross section of semiconducting junctions. 
Indeed, the Majorana condition in these setups is only fulfilled by a few transverse modes~\cite{PhysRevB.86.121103}. Even though a perfect one-dimensional regime is challenging to achieve, most Majorana platforms attempt to approach a one-dimensional limit by reducing the number of transverse modes contributing to transport. 
To study the impact of extra transport modes on the trivial \gls{zbp}, we extend \cref{Eq1} to describe a two-dimensional system and denote as $k_y$ the wavevector component parallel to the interfaces~\cite{SM}. The transport observables in \cref{EqTN,EqdIdV} must then be averaged over all incident modes, and the resonant condition for the confined states in N$_2$ becomes more complicated, as it now depends on the transverse wavevector $k_y$. 
Consequently, as we add extra modes, the magnitude of the \gls{zbp} for trivial junctions on resonance is reduced, although the peak never disappears. 
Even though a quantization of the \gls{zbp} in planar junctions is no longer possible, a conductance approaching the quantized value is still achievable in quasi-one dimensional trivial junctions, see Ref.~\cite{SM}. 

{\it Conclusions.---}We have shown that quasi zero-energy Andreev states can naturally emerge due to confinement effects in hybrid junctions based on conventional $s$-wave superconductors in the absence of any magnetic order. Such confinement-induced  states can mimic the behavior of Majorana zero modes, featuring zero-bias conductance peaks and enhanced zero-energy local density of states. 
Confinement, which here emerges from a depleted or gated finite-length intermediate region, is a very common effect in hybrid junctions, including most Majorana nanowire experiments. Our results thus exemplify how ubiquitous trivial zero-bias peaks can be in hybrid junctions, even without spin fields, and highlight that zero-bias conductance peaks cannot be solely taken as a definitive indicator of Majorana bound states or topological superconductivity. 


We thank  A. M. Black-Schaffer for insightful discussions. 
J. C. acknowledges support from C.F. Liljewalchs stipendiestiftelse Foundation, the Knut and Alice Wallenberg Foundation through the Wallenberg Academy Fellows program, and the EU-COST Action CA-16218 Nanocohybri. 
P. B. acknowledges support from the Spanish CM ``Talento Program'' No. 2019-T1/IND-14088. 

%


%

\clearpage
\onecolumngrid
\setcounter{equation}{0}
\renewcommand{\theequation}{S\,\arabic{equation}}
\setcounter{figure}{0}
\renewcommand{\thefigure}{S\,\arabic{figure}}
\section{Supplemental Material for ``Confinement-induced zero-bias peaks in conventional superconductor hybrids"}

In this Supplementary Material we provide details on how we calculate the conductance, the Green's functions, and the \gls{dos} for N$_{1}$N$_{2}$S junctions. We also detail the numerical simulations for finite-size and quasi-one dimensional systems.

%
\section{Conductance and density of states}
In the main text, we have demonstrated that zero-energy states ubiquitously emerge 
in finite length  junctions due to confinement effects. We have also shown that these zero-energy states give rise to zero-bias conductance peaks and large zero-bias \gls{dos}. To support those findings, here we provide further details on the used models, and on how to obtain conductance and \gls{dos} by using scattering states. 

\subsection{Junction models}
We consider ballistic N$_{1}$N$_{2}$S junctions, where the middle region N$_{2}$ is of finite length $L$, with the left interface  located at $x=-L$ and the right interface at $x=0$. The BdG Hamiltonian presented in the main text for the junction with a conventional $s$-wave spin-singlet superconductor can be written as
\begin{equation}
	\label{BdG2}
	\begin{split}
		H_{\rm BdG}(x)&=\begin{pmatrix}
			H_{0}(x)&\Delta(x)\\
			\Delta^{\dagger}(x)&-H_{0}(x)
		\end{pmatrix}\,,
	\end{split}
\end{equation}
in the basis $\psi_{k}^{\rm T}=(c_{k,\uparrow},c_{-k,\downarrow}^{\dagger})$, 
where $H_{0}(x)=\frac{p^{2}}{2m}-\mu(x)$, with $p=-i\hbar\partial_{x}$. For the considered N$_{1}$N$_{2}$S  junction, the $s$-wave spin-singlet pair potential is given by 
\begin{equation}
	\label{NSDelta}
	\Delta(x)=
	\begin{cases}
		0\,, & x<-L, \\
		0\,, & -L<x<0, \\
		\Delta\,, & x>0\,.
	\end{cases}
\end{equation} 
The chemical potential profile is given by
\begin{equation}
	\label{NSmu}
	\mu(x)=
	\begin{cases}
		\mu_{\rm N_{1}}\,, & x<-L, \\
		\mu_{\rm N_{2}}\,, & -L<x<0, \\
		\mu_{\rm S}\,, & x>0.
	\end{cases}
\end{equation} 
To compare our results with unconventional superconductors, we also consider N$_{1}$N$_{2}$S junctions with S being a topological superconductor. In this case, the modeling is very similar with the difference that S here represents an spinless (spin-polarized) $p$-wave superconductor deep in the topological phase,  as in the topological phase of the well known Kitaev model \cite{kitaev}. Hence, S is modelled by a BdG Hamiltonian similar to Eq.\,(\ref{BdG2}) but now in the spinless basis $\psi_{k}^{\rm T}=(c_{k},c_{-k}^{\dagger})$ and with the order parameter given by $p\Delta/k_{\rm S}$, with $k_{\rm S}$ the Fermi wave vector in S. The N regions are here spin-polarized or spinless, a property shared with the S region as well; this is expected for a strong magnetic field applied along the wire needed to reach the topological phase in the S region based on the conventional realizations of topological superconductivity~\cite{Aguadoreview17}.

\subsection{Scattering states}
\label{AppScatS}

We now discuss how to obtain the conductance and \gls{dos} based on the scattering states for the \gls{bdg} Hamiltonian given by Eq.\,(\ref{BdG2}). 
In order to construct the scattering states, we first solve the eigenvalue problem in each region (N$_{i}$ and S) by using the Hamiltonian given by Eq.\,(\ref{BdG2}). Next, we construct the scattering states taking into account right-moving electrons and holes from the leftmost region (N$_{1}$) and left-moving quasielectrons and quasiholes from the rightmost region (S). The resulting four scattering processes for the N$_{1}$N$_{2}$S junction are given by
\begin{equation}
	\label{normalw}
	\begin{split}
		\Psi_{1}(x)&=
		\begin{cases}
			\phi_{1}^{\rm N}\,{\rm e}^{ik_{e_{1}}x}+a_{1}\phi_{3}^{\rm N}\,{\rm e}^{ik_{h_{1}}x} +b_{1}\phi_{2}^{\rm N}\,{\rm e}^{-ik_{e_{1}}x},\,\quad\quad\,\,\, \quad\quad\quad\quad\quad\quad\quad\quad\quad x<-L&  \\
			e_{1} \phi_{1}^{\rm N}\,{\rm e}^{ik_{e_{2}}x}+  e_{2} \phi_{2}^{\rm N}\,{\rm e}^{-ik_{e_{2}}x}
			+ e_{3} \phi_{3}^{\rm N}\,{\rm e}^{ik_{h_{2}}x}+e_{4} \phi_{4}^{\rm N}\,{\rm e}^{-ik_{h_{2}}x}\,,
			\quad\quad\,\,\,-L<x<0&  \\
			c_{1}  \phi_{1}^{\rm S}\,{\rm e}^{iq_{e}x} +d_{1}  \phi_{4}^{\rm S}\,{\rm e}^{-iq_{h}x},\,\quad \quad \quad \quad \quad \quad \quad\quad\quad \quad\quad\quad\quad\quad\quad\quad\quad\quad\, x>0 & 
		\end{cases} , \\
		\Psi_{2}(x)&=
		\begin{cases}
			\phi_{4}^{\rm N}\,{\rm e}^{-ik_{h_{1}}x}+a_{2}\phi_{2}^{\rm N}\,{\rm e}^{-ik_{e_{1}}x} +b_{2}\phi_{3}^{\rm N}\,{\rm e}^{ik_{h_{1}}x},\quad\quad \quad\quad\quad\quad\quad\quad\quad\quad\quad x<-L& \\
			f_{1} \phi_{1}^{\rm N}\,{\rm e}^{ik_{e_{2}}x}+  f_{2} \phi_{2}^{\rm N}\,{\rm e}^{-ik_{e_{2}}x}
			+ f_{3} \phi_{3}^{\rm N}\,{\rm e}^{ik_{h_{2}}x}+f_{4} \phi_{4}^{\rm N}\,{\rm e}^{-ik_{h_{2}}x}\,,      
			\quad\quad\,\,-L<x<0&  \\
			c_{2}  \phi_{4}^{\rm S}\,{\rm e}^{-iq_{h}x} +    d_{2}  \phi_{1}^{\rm S}\,{\rm e}^{iq_{e}x}, \,\quad \quad \quad \quad \quad \quad \quad \quad \quad \quad\quad\quad\quad\quad\quad\quad\quad\quad\, x>0 & 
		\end{cases} , \\
		\Psi_{3}(x)&=
		\begin{cases}
			c_{3}  \phi_{2}^{\rm N}\,{\rm e}^{-ik_{e_{1}}x} +d_{3}  \phi_{3}^{\rm N}\,{\rm e}^{ik_{h_{1}}x},\,\quad \quad \quad \quad \quad\quad\quad\quad\quad\quad\quad\quad\quad \quad\quad\quad\,  x<-L&\\
			g_{1} \phi_{1}^{\rm N}\,{\rm e}^{ik_{e_{2}}x}+  g_{2} \phi_{2}^{\rm N}\,{\rm e}^{-ik_{e_{2}}x}
			+ g_{3} \phi_{3}^{\rm N}\,{\rm e}^{ik_{h_{2}}x}+g_{4} \phi_{4}^{\rm N}\,{\rm e}^{-ik_{h_{2}}x}\,,  
			\quad\quad\,\,
			-L<x<0&  \\
			\phi_{2}^{\rm S}\,{\rm e}^{-iq_{e}x}+a_{3}\phi_{4}^{\rm S}\,{\rm e}^{-iq_{h}x}+b_{3}\phi_{1}^{\rm S}\,{\rm e}^{iq_{e}x},\,\quad\quad\quad\quad \quad\quad\quad\quad\quad\quad\quad\quad\quad x>0&  
		\end{cases} , \\
		\Psi_{4}(x)&=
		\begin{cases}
			c_{4}  \phi_{3}^{\rm N}\,{\rm e}^{ik_{h_{1}}x} +    d_{4}  \phi_{2}^{\rm N}\,{\rm e}^{-ik_{e_{1}}x},\,\quad \quad \quad \quad \quad\quad\quad\quad\quad\quad\quad\quad\quad \quad\quad\quad\, x<-L & \\
			h_{1} \phi_{1}^{\rm N}\,{\rm e}^{ik_{e_{2}}x}+  h_{2} \phi_{2}^{\rm N}\,{\rm e}^{-ik_{e_{2}}x}
			+ h_{3} \phi_{3}^{\rm N}\,{\rm e}^{ik_{h_{2}}x}+h_{4} \phi_{4}^{\rm N}\,{\rm e}^{-ik_{h_{2}}x}\,,
			\quad\quad-L<x<0&  \\
			\phi_{3}^{\rm S}\,{\rm e}^{iq_{h}x}+a_{4}\phi_{1}^{\rm S}\,{\rm e}^{iq_{e}x}+b_{4}\phi_{4}^{\rm S}\,{\rm e}^{-iq_{h}x}\,,\quad\quad\quad\quad\quad\quad\quad\quad\quad\quad\quad \quad\quad\,\,\,  x>0&
		\end{cases} , 
	\end{split}
\end{equation}
where 
\begin{equation}
	\begin{split}
		\phi_{1,2}^{\rm N}&=
		\begin{pmatrix}
			1\\
			0
		\end{pmatrix} , \quad 
		\phi_{3,4}^{\rm N}=
		\begin{pmatrix}
			0\\
			1
		\end{pmatrix} , \quad 
		\phi_{1,2}^{\rm S} =
		\begin{pmatrix}
			u\\
			v
		\end{pmatrix} , \quad 
		\phi_{3,4}^{\rm S}=
		\begin{pmatrix}
			\eta v\\
			u
		\end{pmatrix},
	\end{split}
\end{equation}
are the spinors of the N$_{i}$ and S regions, with $\eta=+$ ($\eta=-$) for an $s$-wave ($p$-wave) superconductor. We have used the superconducting coherence factors
\begin{equation}
	\begin{split}
		u=\sqrt{\frac{1}{2}\left(1+\frac{\sqrt{\omega^{2}+\Delta^{2}}}{\omega}\right)},\quad
		v=\sqrt{\frac{1}{2}\left( 1-\frac{\sqrt{\omega^{2}+\Delta^{2}}}{\omega}\right)},
	\end{split}
\end{equation}
and 
\begin{equation}
	\begin{split}
		k_{e_{i}(h_{i})}= \sqrt{\frac{2m}{\hbar^{2}} \left(\mu_{\rm N_{i}}\pm \omega \right)}\,,\quad
		q_{e(h)}=\sqrt{\frac{2m}{\hbar^{2}}\left( \mu_{\rm S}\pm \sqrt{\omega^{2}-\Delta^{2}}\right)}\,,\\
	\end{split}
\end{equation}
representing the electron (hole) wavevectors in the normal  N$_{1}$ and superconducting S regions, respectively. 

The four scattering processes  $\Psi_{i}$ in Eqs.\,(\ref{normalw}) represent, respectively, incident right-moving electron, right-moving hole, left-moving quasi-electron, and left-moving quasi-hole. The coefficients of the scattering states have a very specific meaning, associated to the nature of scattering process they describe. For instance, for $\Psi_{1}$, $a_{1}$ and $b_{1}$ are the amplitude of Andreev and normal reflections in N$_{1}$, respectively. Similarly, $c_{1}$ and $d_{1}$ are the amplitude of transmission into a quasielectron and quasihole, respectively. 

In order to fully determine the scattering states $\Psi_{i}$ we still need to calculate their coefficients $a_{i}$, $b_{i}$, $c_{i}$, $d_{i}$, $d_{i}$, $e_{i}$, $f_{i}$, $g_{i}$, and $h_{i}$, with $i=1,2,3,4$. This is carried out by  using the conditions established when integrating the BdG equations at both interfaces, which read
\begin{equation}
	\begin{split}
		\label{conditions}
		\Big[\partial_{x}\Psi(0>x>-L_{\rm N})] -[\partial_{x}\Psi(x<-L_{\rm N})]&=0\,,\\
		\Big[\partial_{x}\Psi(x>0)] -[\partial_{x}\Psi(x<0)]&=0\,,\\
		[\Psi_{i}(x<-L_{\rm N}))]&=[\Psi_{i}(0>x>-L_{\rm N}))]\,,\\
		[\Psi_{i}(x<0)]&=[\Psi_{i}(x>0)]\,.
	\end{split}
\end{equation}
This way we obtain all the coefficients of the scattering states in Eqs.\,(\ref{normalw}),
for N$_{1}$N$_{2}$S junctions with S being either $s$-wave or $p$-wave superconductor deep in the topological phase. Here, junctions with an $s$-wave superconductor are referred to as trivial junctions, while those with a $p$-wave superconductor in the topological phase are referred to as topological junctions. 

The coefficients of Eqs.\,(\ref{normalw}) allow us to obtain the local conductance in N$_{1}$ at zero temperature as,
\begin{equation}
	\label{conductance}
	\sigma(\omega)=\frac{2e^{2}}{h}\big[1-|b_{1}|^{2}+|a_{1}|^{2}\big]\,,
\end{equation}
where $a$ and $b$ represent the amplitude of Andreev and normal reflection amplitudes, respectively. Note that Eq.\,(\ref{conductance}) is valid for both trivial and topological junctions. 
\cref{conductance} is used to obtain the conductance in Fig.\,2 of the main text for trivial and topological junctions. 

In the case of a trivial ($\eta=+$) and a topological ($\eta=-$) junction, the normal and Andreev reflection amplitudes are
\begin{equation}
	\begin{split}
		a_{1}={}& -\frac{8 \Gamma}{t} k_{e1} k_{e2} k_{h2} \left(q_e+q_h\right)\e^{-i(k_{e1} - k_{h1} - k_{e2} - k_{h2})L} ,\\
		b_{1}={}& -\frac{1}{t} \left\{
		\left(k_{e1} + k_{e2}\right) \left(k_{h1} - k_{h2}\right) \left[ \left(q_{e} - k_{e2}\right) \left(q_{h} - k_{h2}\right) - \eta\Gamma^2 \left(q_{h} + k_{e2}\right) \left(q_{e} + k_{h2}\right) \right] \right. \\
		&\left. + \left(k_{e1} - k_{e2}\right) \left(k_{h1} + k_{h2}\right) \left[ \left(q_{e} + k_{e2}\right) \left(q_{h} + k_{h2}\right) - \eta\Gamma^2 \left(q_{h} - k_{e2}\right) \left(q_{e} - k_{h2}\right) \right] \e^{-2i (k_{e2}-k_{h2})L} \right. \\
		&\left. - \left(k_{e1} - k_{e2}\right) \left(k_{h1} - k_{h2}\right) \left[ \left(q_{e} + k_{e2}\right) \left(q_{h} - k_{h2}\right) - \eta\Gamma^2 \left(q_{h} - k_{e2}\right) \left(q_{e} + k_{h2}\right) \right] \e^{-2i k_{e2}L} \right. \\
		&\left. - \left(k_{e1} + k_{e2}\right) \left(k_{h1} + k_{h2}\right) \left[ \left(q_{e} - k_{e2}\right) \left(q_{h} + k_{h2}\right) - \eta\Gamma^2 \left(q_{h} + k_{e2}\right) \left(q_{e} - k_{h2}\right) \right] \e^{2i k_{h2}L} \right\},\\
		t={}& 
		\left(k_{e1} + k_{e2}\right) \left(k_{h1} - k_{h2}\right) \left[ \left(q_{e} + k_{e2}\right) \left(q_{h} - k_{h2}\right) - \eta\Gamma^2 \left(q_{h} - k_{e2}\right) \left(q_{e} + k_{h2}\right) \right] \\
		+&\left(k_{e1} - k_{e2}\right) \left(k_{h1} + k_{h2}\right) \left[ \left(q_{e} - k_{e2}\right) \left(q_{h} + k_{h2}\right) - \eta\Gamma^2 \left(q_{h} + k_{e2}\right) \left(q_{e} - k_{h2}\right) \right] \e^{2i (k_{e2}+k_{h2})L} \\
		-&\left(k_{e1} - k_{e2}\right) \left(k_{h1} - k_{h2}\right) \left[ \left(q_{e} - k_{e2}\right) \left(q_{h} - k_{h2}\right) - \eta\Gamma^2 \left(q_{h} + k_{e2}\right) \left(q_{e} + k_{h2}\right) \right] \e^{2i k_{e2}L} \\
		-&\left(k_{e1} + k_{e2}\right) \left(k_{h1} + k_{h2}\right) \left[ \left(q_{e} + k_{e2}\right) \left(q_{h} + k_{h2}\right) - \eta\Gamma^2 \left(q_{h} - k_{e2}\right) \left(q_{e} - k_{h2}\right) \right] \e^{2i k_{h2}L}  ,
	\end{split}
\end{equation}
with $\Gamma=|\Delta|/(\omega+\sqrt{\omega^2-\Delta^2})$.

\begin{figure}[!t]
	\centering
	\includegraphics[width=.7\textwidth]{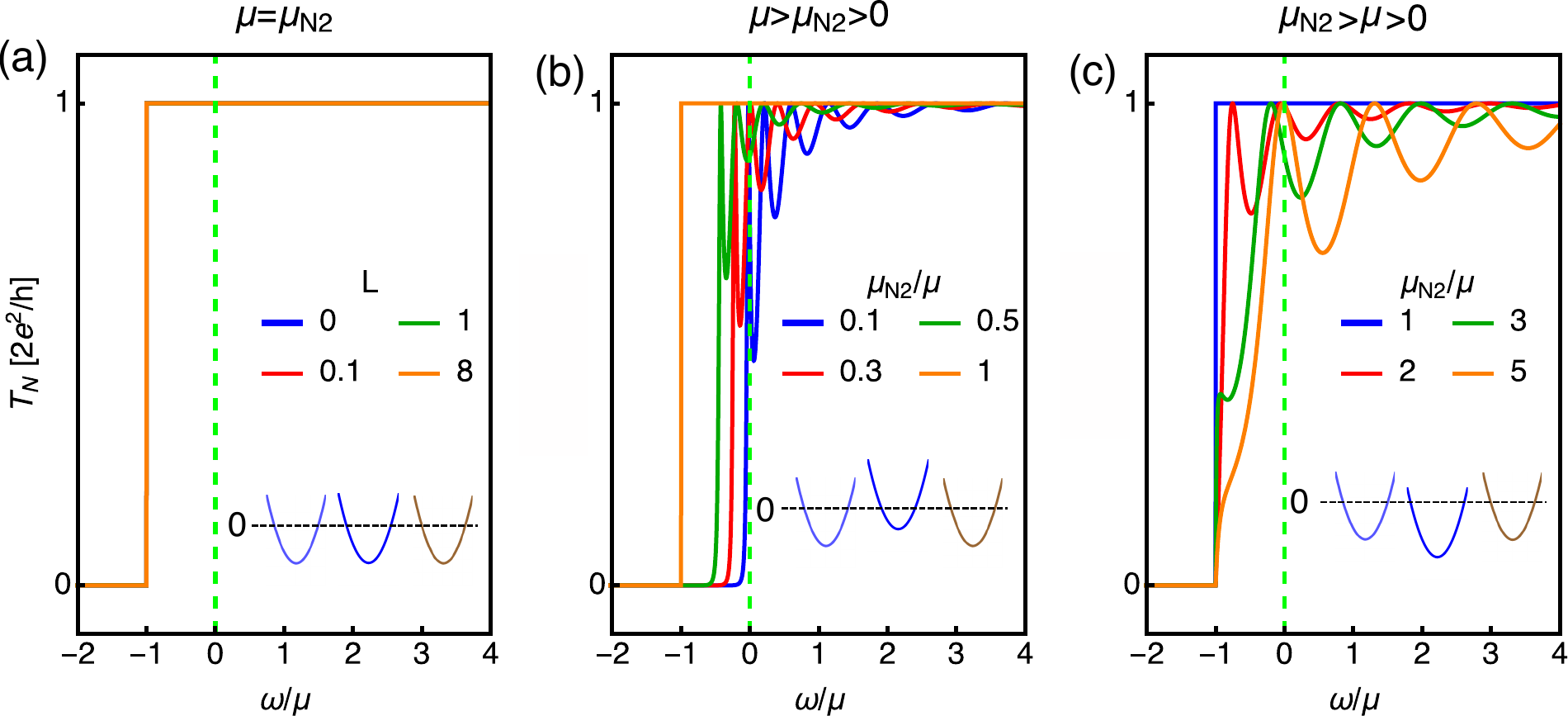} 
	\caption{Conductance of an N$_{1}$N$_{2}$S junction when S is in the normal state, for (a) equal chemical potentials $\mu=\mu_{\rm N_{2}}>0$, (b) $\mu>\mu_{\rm N_{2}}>0$, and (c) $0<\mu<\mu_{\rm N_{2}}$. The insets depict the position of the bands under the given chemical potentials with respect to zero energy.}
	\label{fig:TN}
\end{figure}

When superconductivity vanishes, and the leftmost and rightmost regions are normal metals with the same chemical potential, the conductance adopts a simplified form given by
\begin{equation}
	\label{TN}
	T_{\rm N}(\omega)=\frac{2e^{2}}{h}\frac{1}{1+\big[\frac{k_{e1}}{2k_{e2}}+\frac{k_{e2}}{2k_{e1}}\big]^{2}-\big[\frac{k_{e1}}{2k_{e2}}-\frac{k_{e2}}{2k_{e1}}\big]^{2}{\rm cos}(2k_{e_{2}}L)}\,.
\end{equation}
where $k_{e1}$ is the electron wave vector for both the leftmost and rightmost regions, while $k_{e_{2}}$ belongs to the middle one. Equation~(\ref{TN}) results from the following scattering processes: 
an incident electron from N$_{1}$ can be transmitted through N$_2$ directly into the rightmost region N$_3$, or it can be trapped inside the finite region N$_{2}$, experience several reflections at both interfaces, and get finally transmitted to $N_{3}$. 
Consequently, the finite length region N$_{2}$ behaves like a cavity, trapping electrons which develop an interference pattern because the incident electronic state has two possible paths through N$_{2}$. This effect is reflected in the denominator of $T_{\rm N}$ in the form of an oscillatory cosine function that depends on the properties of the finite middle region N$_{2}$. 
The cosine term thus permits the appearance of discrete energy levels in the finite region. The oscillatory modulation disappears either when $L=0$ (no middle region or an extremely short one), or when we have the same wave vectors in all the three regions, i.e., $k_{e1}=k_{e2}$. The latter case requires the same chemical potentials in all three regions. Whenever one of these two conditions are satisfied we get a normal transmission of $T_{\rm N}=1$, an instance of resonant transport as seen in Fig.\,\ref{fig:TN}(a); see also orange and blue curves in (b) and (c), respectively. 

Interestingly, the normal-state transmission $T_{\rm N}$ features an energy modulation when the middle region is finite and its chemical potential different from that of the outer regions. 
The energy modulation can be enhanced when, e.g., we take the limit $0<k_{e1}\ll2k_{e2}$, where the transmission acquires the following form $T_{\rm N}\approx2/[1+(k_{e1}^{2}/2k_{e2}^{2}){\rm sin}^{2}(k_{e2}L)]<1$. Analogously, for $0<k_{e2}\ll2k_{e1}$, we get $T_{\rm N}\approx2/[1+(k_{e2}^{2}/2k_{e1}^{2}){\rm sin}^{2}(k_{e2}L)]<1$. 
In these two regimes the transmission is reduced to $T_{\rm N}\ll1$ when the sine term has maximal amplitude ($\sin^2(k_{e2}L)\sim1$). Figure~\ref{fig:TN} shows how the oscillatory modulation tends to reduce the values of the perfect transmission as a result of the phases acquired by the electrons inside the finite region. 
This effect coincides with the appearance of discrete energy levels inside N$_{2}$, which can be seen as resonances of the transmission in Fig.\,\ref{fig:TN} (b,c). An interesting feature to note in Fig.\,\ref{fig:TN} (b,c) is that $T_{\rm N}$ can be peaked at zero energy $\omega=0$, solely as a result of the finite length and different chemical potentials. We show in the main text that this innocent behavior has interesting consequences in the physics of hybrid junctions.

The local conductance in the normal state plotted in \cref{fig:TN} is presented in Eq.\,(2) of the main text. 

\subsection{Retarded Green's functions}
We now detail how to obtain the \gls{dos} presented in Eq.\,(6) and Fig.\,3 of the main text. To calculate the \gls{dos} we employ the retarded Green's function of the system. Thus, we first construct the retarded Green's function $G^{r}(x,x',\omega)$ with outgoing boundary conditions in each region from the scattering processes at the interfaces~\cite{PhysRev.175.559}. Thus, in general, the retarded Green's function can be calculated as~\cite{PhysRev.175.559,Kashiwaya_2000} 
\begin{equation}
	\label{RGF}
	\begin{split}
		G^{r}(x,x',\omega)=
		&\begin{cases}
			\alpha_{1} \Psi_{1}(x)\tilde{\Psi}_{3}^{T}(x')+\alpha_{2} \Psi_{1}(x)\tilde{\Psi}_{4}^{T}(x') 
			+ \alpha_{3} \Psi_{2}(x)\tilde{\Psi}_{3}^{T}(x')+\alpha_{4} \Psi_{2}(x)\tilde{\Psi}_{4}^{T}(x'), & x>x', \\
			\beta_{1} \Psi_{3}(x)\tilde{\Psi}_{1}^{T}(x')+  \beta_{2} \Psi_{4}(x)\tilde{\Psi}_{1}^{T}(x') 
			+   \beta_{3} \Psi_{3}(x)\tilde{\Psi}_{2}^{T}(x')+  \beta_{4} \Psi_{4}(x)\tilde{\Psi}_{2}^{T}(x'), & x<x',
		\end{cases}
	\end{split}
\end{equation}
where $\Psi_{i}$ represent the scattering processes at the interfaces of the junction under investigation (N$_{1}$N$_{2}$S) and  given by Eqs.\,(\ref{normalw}). Here, $\tilde{\Psi}_{i}$ corresponds to the conjugated scattering processes obtained using $\tilde{H}_{\rm BdG}(k)=H_{\rm BdG}^{*}(-k)=H_{\rm BdG}^{\rm T}(-k)$ instead of Eq.\,(\ref{BdG2}). In general, the component of $\Psi_{i}$ (either for N$_{i}$ or S) used will determine the range of $x$ and $x'$.

The coefficients $\alpha_{i}$ and $\beta_{i}$ in Eq.~\eqref{RGF} are found from the continuity of the Green's function
\begin{equation}
	[\omega-H_{\rm BdG}(x)]G^{r}(x,x',\omega)=\delta(x-x')\,,
\end{equation}
where $H_{\rm BdG}$ is the BdG Hamiltonian of the system defined in Eq.\,(\ref{BdG2}). Then, by integrating around $x=x'$ we obtain
\begin{equation}
	\label{conditionGR}
	\begin{split}
		&[G^{r}(x>x')]_{x=x'}=[G^{r}(x<x')]_{x=x'}\,,\\
		&[\partial_{x}G^{r}(x>x')]_{x=x'}-[\partial_{x}G^{r}(x<x')]_{x=x'}=(2m/\hbar^{2})\tau_{z}\,,
	\end{split}
\end{equation}
where $\tau_{i}$ are $i$-Pauli matrices in electron-hole space. 

In general, the Green's function, either in  N$_{i}$ or S region, is a $2\times2$ matrix in electron-hole space,
\begin{equation}
	\label{GF}
	G^{r}(x,x',\omega)=
	\begin{pmatrix}
		G^{r}_{0}(x,x',\omega)&F(x,x',\omega)\\
		\bar{F}(x,x',\omega)&\bar{G}^{r}_{0}(x,x',\omega)
	\end{pmatrix}\,,
\end{equation}
where $G^{r}_{0}$ and $F$ correspond to the normal and anomalous components, respectively. Note that each element in the previous matrix is a scalar number because spin is not active. The diagonal term $G_{0}^{r}$ is particularly relevant because it allows us to calculate the \gls{dos} as $\rho(\omega,x)=(1/\pi){\rm Im}{\rm Tr}\big[G^{r}(x,x,\omega)\big]$. 

\begin{figure*}[!ht]
	\begin{minipage}[t]{\linewidth}
		\centering
		\includegraphics[width=0.75\columnwidth]{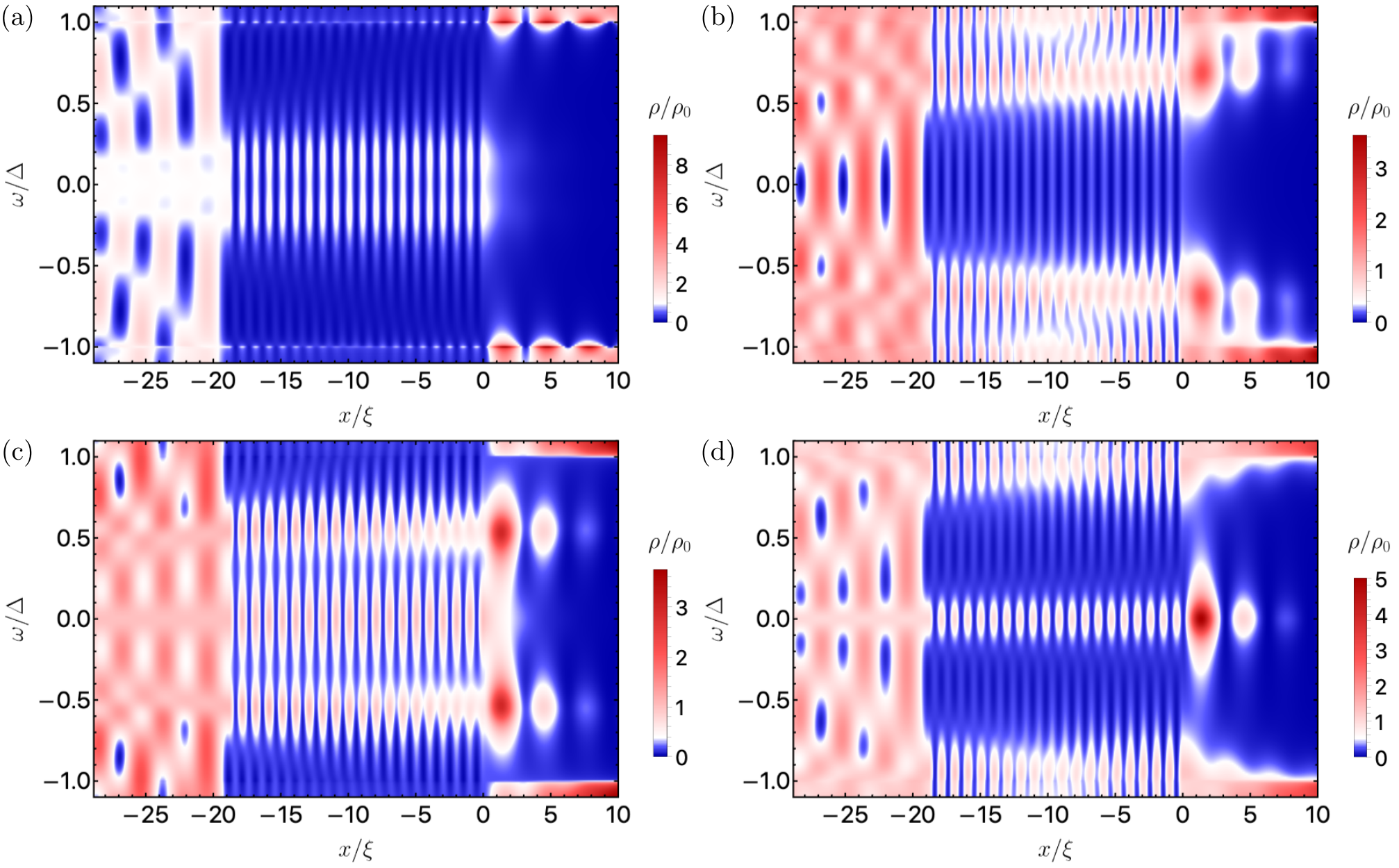}
		\caption{Map of the \gls{dos} as a function of the energy and the position for trivial (a,b) and topological (c,d) superconductors. Left (a,c) and right (b,d) panels correspond to on- and off-resonance cases, respectively, with values of $\mu_{\rm N_{2}}$ indicated by the dashed gray lines in Fig. 3(a) of the main text.  All parameters are the same as in Fig. 3 of the main text.  }
		\label{fig:DOS}
	\end{minipage}
\end{figure*}

We follow the approach outlined above  and calculate the Green's functions  to obtain the \gls{dos} for trivial and topological junctions. \Cref{fig:DOS} shows the \gls{dos} as a function of energy and space when the chemical potential of the middle region is set on-resonance [\cref{fig:DOS} (a,c)] and off-resonance [\cref{fig:DOS}(b,d)]; line cuts of these density plots are presented in Fig.\,3 of the main text.  In the N$_{1}$ region it is possible to find a simple expression which is then given by Eq.\,(4) in the main text.

On-resonance, the zero-energy \gls{dos} for a trivial junction behaves very similar as for a topological junction, as seen in Fig.\,(\ref{fig:DOS}). This, however, breaks down off-resonance, where the  zero-energy \gls{dos} remain robust for the topological junction but is drastically reduced for the trivial junctions. While there are clear differences, these results point out that it might be challenging to distinguish the trivial and topological zero-energy \gls{dos} on-resonance. 

\begin{figure*}[!ht]
	\begin{minipage}[t]{\linewidth}
		\centering
		\includegraphics[width=0.75\columnwidth]{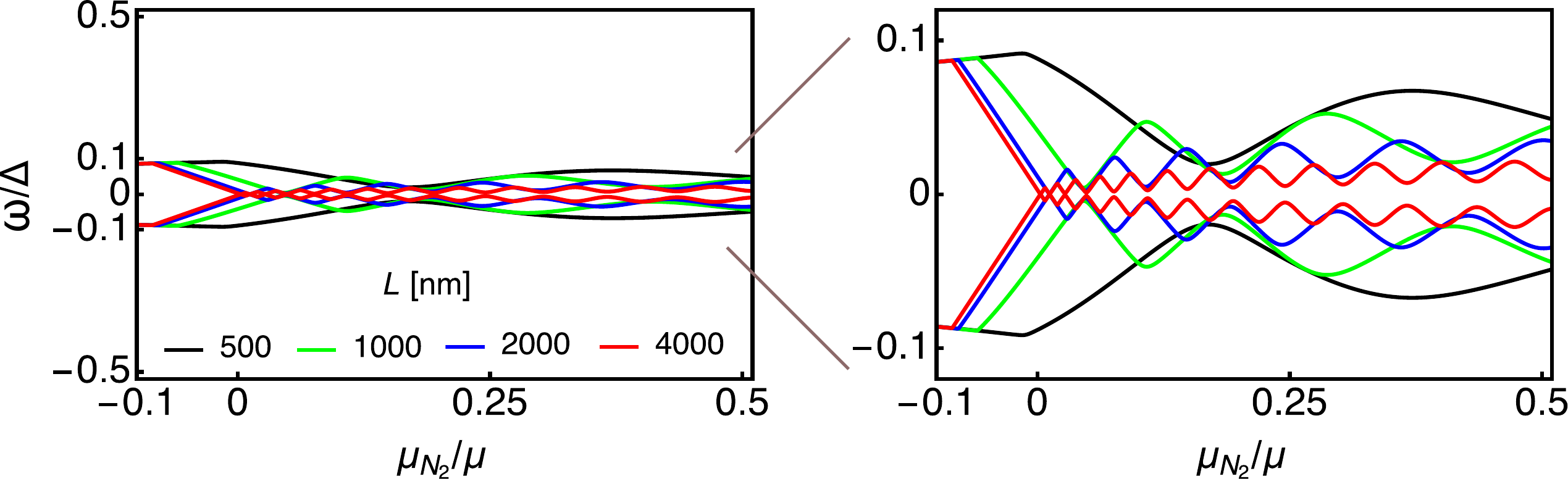}
		\caption{Lowest energy levels of a trivial N$_{1}$N$_{2}$S with all regions being of finite length. Parameters: $L_{\rm N_{1}}=L_{\rm S}=2000$\,nm, $\Delta=0.5$\,meV, $\mu=0.5$\,meV.}
		\label{fig:TBqZESs}
	\end{minipage}
\end{figure*}

\section{Quasi zero-energy states in a finite length junction}
In this part we show that the trivial quasi zero-energy states, discussed in the main text, remain even when the leftmost and rightmost regions have finite length, which we denote, respectively, as $L_{\rm N_{1}}$ and $L_{\rm S}$. As in the previous sections, the width of the central region is denoted $L$. 
We thus discretize Eq.\,(\ref{BdG2}) into a tight-binding lattice with a lattice spacing of $a=10$~nm and inspect the lowest energy level of the spectrum. This calculation is presented in \cref{fig:TBqZESs} for realistic parameters as a function of the chemical potential in the middle region $\mu_{\rm N_{2}}$, for different values of $L$. 
We observe that the lowest energy levels oscillate with $\mu_{\rm N_{2}}$ and approach zero-energy for sufficiently long middle regions. 
While the lowest energy states do not truly pin to zero energy, when they approach on resonance the gap between them is much smaller than one tenth of the superconducting gap. Such a small splitting would be challenging to resolve experimentally. 
We believe this is the critical issue when interpreting conductance data as, for all practical purposes, the \textit{trivial} lowest energy states behave as a pair of Majorana bound states. 
We conclude this part by saying that the quasi zero-energy states presented in the main text remain robust even in finite length junctions and, thus, can be expected to occur in realistic setups.

\begin{figure*}[!ht]
	\begin{minipage}[t]{\linewidth}
		\centering
		\includegraphics[width=0.55\columnwidth]{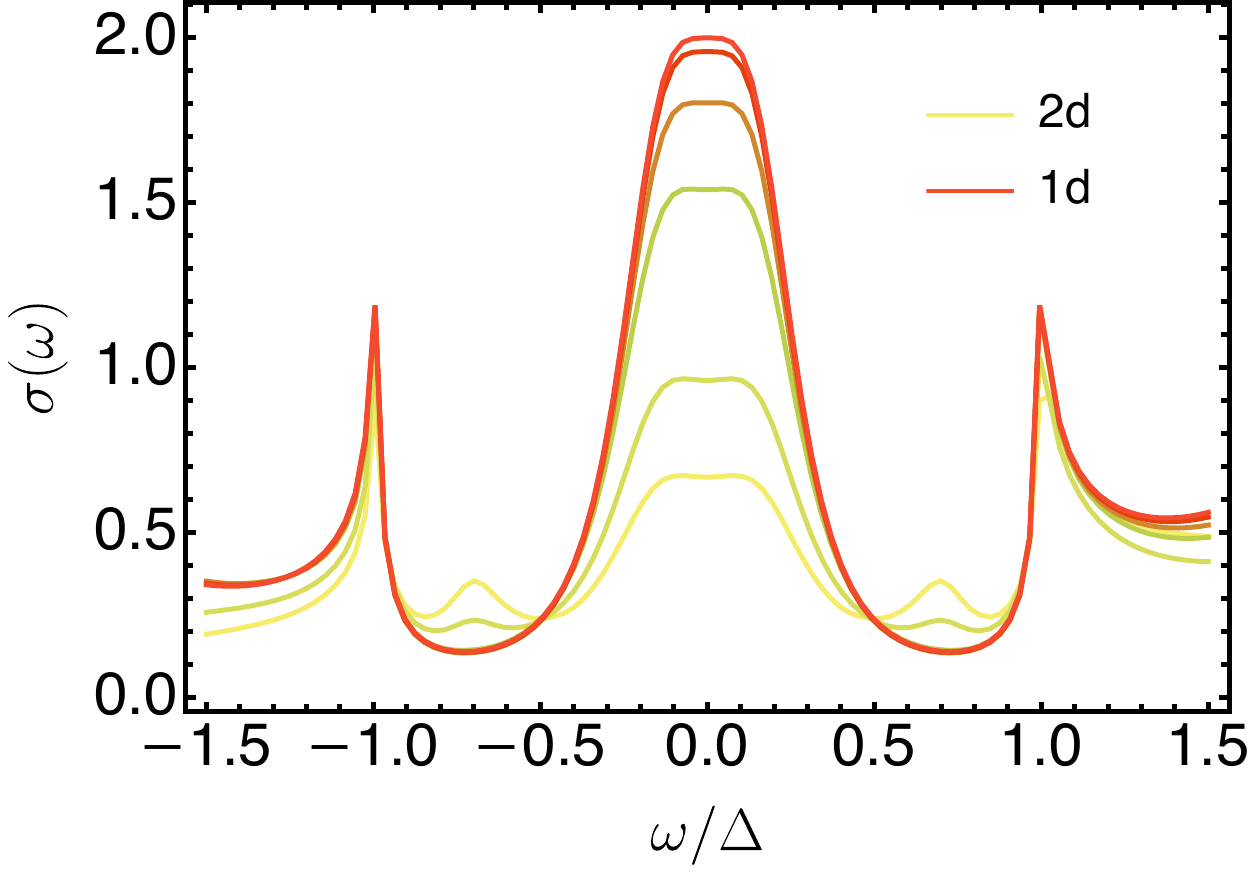}
		\caption{Conductance of the N$_1$N$_2$S junction in the quasi-one dimensional limit. The width and doping of the central region is set on resonance with $\mu_{\text{N}2}=9.03\mu$, $k_\text{F}L=3\pi/2$, and $\mu_{\text{N}1}=\mu_\text{S}\equiv\mu=2\Delta$. 
			We set $\lambda=0$ (two-dimensional limit), $\lambda=2,4,6,10$, and $\lambda=20$ (equivalent to one-dimensional case). 
		}
		\label{fig:quasi1d}
	\end{minipage}
\end{figure*}

\section{Quasi-one dimensional junctions}
In the main text, we only considered one dimensional junctions. Most platforms for implementing Majorana bound states, even not being perfectly one-dimensional, only approach this limit after reducing the number of transverse modes contributing to transport. 
We now go beyond the one-dimensional limit expanding the Hamiltonian of Eq. (1) into the quasi-one dimensional limit. As a result, the momentum operator is written as $p=-i\hbar(\partial_x+\partial_y)$. Considering that transport takes place along the $x$ direction, and each interface is perfectly flat in the $y$ direction, we can parametrize the transverse component of the wave vector by the angle of incidence on the leftmost normal region, $\theta=\sin^{-1}(k_y/k_{\text{N}1})$. For the other regions, translational invariance along the $y$ direction entails that 
\begin{equation}
	k_{\text{N}1}\sin\theta = k_{\text{N}2}\sin\theta_{\text{N}2} = k_{\text{S}}\sin\theta_\text{S} .
\end{equation}
Consequently, the conductance in the normal and superconducting states, Eq. (2) and Eq. (3), respectively, must be averaged over all incident modes as
\begin{equation}\label{eq:angle-average}
	\mean{\sigma(\omega)}_{\mbf{k}_\parallel} = \frac{1}{2} \int\limits^{\pi/2}_{\pi/2}  \mathrm{d}\theta \cos\theta P(\theta) \sigma(\omega). 
\end{equation}
Here, $P(\theta)$ is the probability distribution for the transverse modes at the interface. We choose a Gaussian function of the form
\begin{equation}
	P(\theta)= \exp\left( - \frac{\lambda^2}{2} \theta^2 \right), 
\end{equation}
which allows us to recover the one-dimensional limit for $\lambda\rightarrow\infty$, while the planar two-dimensional case corresponds to $\lambda\rightarrow0$. In the former case, we recover the results of the main text, while the latter case is equivalent to setting $P(\theta)=1$ in \cref{eq:angle-average}. 

The effect of adding extra transverse modes to the one-dimensional conductance is shown in \cref{fig:quasi1d}. To study the impact on the trivial zero energy states, we chose a junction on resonance featuring a quantized zero-bias conductance peak. 
First, the one-dimensional limit, red line in \cref{fig:quasi1d}, is already reproduced for $\lambda=20$. 
Reducing the value of the parameter $\lambda$, we approach the fully two-dimensional limit of a planar junction when $\lambda=0$ (yellow line). As we add extra modes, the magnitude of the zero-bias conductance peak is reduced, although the peak never disappears. The values in \cref{fig:quasi1d} correspond, in descending order from the one-dimensional case, to $\lambda=10, 6, 4$, and $2$.

\end{document}